\documentclass[aps, pra, twocolumn, notitlepage, superscriptaddress]{revtex4-1}
\usepackage{graphicx}
\usepackage{amsmath}
\usepackage{amssymb}
\usepackage{color}
\usepackage{multirow}
\usepackage[english]{babel}
\usepackage{hyperref}

\begin{document}
\title{Valley polarized current and resonant electronic transport\\ in a nonuniform  $\mathbf{MoS}_2$ zigzag nanoribbon}

\author{D. Gut}
\affiliation{AGH University of Science and Technology, Faculty of Materials Science and Ceramics, al. A. Mickiewicza 30, 30-059 Krakow, Poland}
\author{M. Prokop}
\affiliation{AGH University of Science and Technology, Faculty of Metals Engineering and Industrial Computer Science, al. A. Mickiewicza 30, 30-059 Krakow, Poland}
\author{D. Sticlet}
\affiliation{National Institute for Research and Development of Isotopic and Molecular Technologies, 67-103 Donat, 400293 Cluj-Napoca, Romania}
\author{M. P. Nowak}
\affiliation{AGH University of Science and Technology, Academic Centre for Materials and Nanotechnology, al. A. Mickiewicza 30, 30-059 Krakow, Poland}

\begin{abstract}
Using the tight-binding approach we study the electronic transport in a $\mathrm{MoS}_2$ zigzag ribbon with a spatially varying potential profile. Considering a ribbon with a smooth potential step in the Fermi energy regime where the transport is dominated by the edge modes, we find that the conductance exhibits sharp resonances due to the resonant transport through a n-p-n junction effectively created in the structure. We show that in a gated wire the current carried on the wire edges can be blocked despite the metallic band structure of the ribbon. For the Fermi energies corresponding to $\mathrm{MoS}_2$ bulk conduction band, we identify states of the semi-infinite wire that are polarized in the $K$, $K'$, $Q$ valleys and exhibit the valley Hall effect distinctly visible in a nonuniform ribbon. Finally, we show that well-defined momenta of the valley polarized modes allow nearly complete valley polarization of the current in a locally gated ribbon. 
\end{abstract}

\maketitle

\section{Introduction}
In recent years, after the discovery of graphene, there has been increasing interest in two-dimensional materials~\cite{butler_progress_2013, geim_van_2013}. Within this group  semiconducting layers of transition metal dichalcogenides (TMDCs)~\cite{wang_electronics_2012} are particularly attracting. TMDCs are materials with the formula $\mathrm{MX_2}$ where M is a transition metal from the group VI (Mo, W, etc.) and X is a chalcogen (S, Se, Te). Their properties have been analyzed for over 50 years~\cite{wilson_transition_1969} but focused initially on bulk materials.
Only later it was realized that the change in the number of the atomic layers significantly alters the band structure~\cite{mak_atomically_2010} of TMDCs. Especially, the transition from an indirect to a direct band gap of 1.5-2 eV is observed~\cite{kuc_influence_2011, wang_electronics_2012} when the material becomes a monolayer.
The presence of the direct band gap within the optical range of the energy spectrum makes monolayer TMDCs very promising materials for applications in optoelectronics~\cite{wang_electronics_2012, lopez-sanchez_ultrasensitive_2013}.

One of the most exploited representatives of TMDCs is the molybdenum disulfide ($\mathrm{MoS_2}$) monolayer, which is a sheet of molybdenum atoms in between two layers of sulfur. Recently, impressive advances on $\mathrm{MoS_2}$ samples preparation have been reported, including nanowire fabrication using bottom-up~\cite{li_vapourliquidsolid_2018, yang_deriving_2019}, top-down~\cite{liu_top-down_2013}, and etching~\cite{kotekar-patil_coulomb_2019,kotekar-patil_single_2019} techniques. Most importantly, this progress allowed for a creation of elementary nanodevices that enable studies of quantum transport. Already the electronic transport in gated TMDC monolayers~\cite{radisavljevic_single-layer_2011, kim_genuine_2019, zheng_patterning_2019} or the first measurements of the conductance quantization in the split-gate quantum point contact devices~\cite{sharma_split-gated_2017, marinov_resolving_2017} were reported. 

The physical properties of nanoscale TMDCs devices are sensitive to the termination of the structure. There are two main types of the edge termination: zigzag and armchair, which result in the appearance of gapless and gapped edge modes, respectively. The presence of edge modes was demonstrated experimentally in single layer small TMDCs islands~\cite{zhang_direct_2014} with the zigzag edge~\cite{bollinger_one-dimensional_2001, bollinger_atomic_2003, li_mos2_2008} which are typical when $\mathrm{MoS_2}$ is tailored into a nanoribbon~\cite{koos_stm_2016, wang_mixed_2010, liu_top-down_2013}. 

Similar to graphene, the conduction and valence band edges of monolayer TMDCs  are located at the corners of the hexagonal Brillouin zone in the $K$ and $K'$ points. 
The presence of these nonequivalent sets of points gives the charge carriers the pseudospin---valley---degree of freedom~\cite{xu_spin_2014} which can be studied optically~\cite{yao_valley-dependent_2008, yu_valley_2015} and may be used for information encoding and processing~\cite{liu_three-band_2013}. While the valley polarization of the charge carriers was extensively studied for bulk monolayers much less attention was paid to its implications on the electronic transport in nanostructures. It has been shown that the valley polarization can be induced in graphene nanoribbons~\cite{rycerz_valley_2007} or for hole charge carriers in TMDCs nanostructures~\cite{hsieh_electrical_2018} by tailoring the shape and potential profile of the structure. 

The goal of this paper is to explain the transport properties of a nonuniform zigzag $\mathrm{MoS_2}$ nanoribbon in the context of the valley polarization of the charge carriers.
For the description of the nanostructures, where both the bulk material properties and the specific edge termination are important, the common approach is to exploit tight-binding (TB) models that overcome the computational effort of DFT. The TB approach for TMDCs was used already in the studies of electronic transport which focused on the properties of strained~\cite{rostami_theory_2015} and disordered~\cite{ridolfi_electronic_2017} wires or on spin polarization in the presence of electric~\cite{heshmati-moulai_phase_2016} and exchange fields~\cite{khoeini_peculiar_2016}.

In this work, exploiting the TB description of a $\mathrm{MoS_2}$ nanoribbon, we demonstrate that, despite the finite size effects in the ribbon, the valley polarization of the charge carriers is preserved and affects the spatial symmetry of the current flowing in the wire. By considering a ribbon with a varying potential profile, we show that even though the band structure is gapless, the flowing current can be gated out and that the conductance exhibits resonances on quasihole states bound to the edge. Finally, our work shows that the signatures of valley polarization may be seen already in linear response regime, in a two terminal nonuniform structure with a potential step. 

This paper is organized as follows: In Sec.~\ref{sec:model} we introduce the model used in the calculations and describe the considered system. Section~\ref{sec:results} discusses first the valley polarization in a nanoribbon with and without structural disorder, and second, the transport properties of a gated wire. The conclusions are given in Sec.~\ref{sec:summary} followed by an Appendix where we provide details of the adopted model.

\section{Model and theory}
\label{sec:model}
\subsection{Spinfull tight-binding model}
In the present paper we use the real-space model developed by Silva-Guill{\'e}n \emph{et al.}~\cite{silva-guillen_electronic_2016-1} which improves over a common three-band approach~\cite{liu_three-band_2013}. It provides the accurate band structure and also the correct orbital composition of the electronic bands as compared to DFT results.

The Hamiltonian of the system is expressed as:
\begin{equation}
\begin{split}
H = \sum_{i,\mu\nu} \varepsilon_{i,\mu\nu}^M c_{i,\mu}^\dag c_{i,\nu} + \varepsilon_{i,\mu\nu}^S b_{i,\mu}^\dag b_{i,\nu} \\ 
+ \sum_{ij,\mu\nu} (t_{ij, \mu\nu}^{MM} c_{i,\mu}^\dag c_{j,\nu} + t_{ij, \mu\nu}^{SS} b_{i,\mu}^\dag b_{j,\nu}) \\ 
+ \sum_{ij,\mu\nu} t_{ij, \mu\nu}^{MS} c_{i,\mu}^\dag b_{j,\nu} + H.c.,
\end{split}
\label{ham_eq}
\end{equation}
where $i,j$ and $\mu,\nu$ run over the lattice sites and atomic orbitals, respectively. $c_{i,μ}$ ($b_{i,ν}$) is the annihilation operator for Mo (S) atoms and $t$ is the hopping coefficient.

The Hilbert space is spanned by a spin-orbital wave function~\cite{silva-guillen_electronic_2016-1}
\begin{equation}
\psi^{\uparrow,\downarrow} = (d_{3z^2-r^2}, d_{x^2-y^2}, d_{xy}, p^S_x, p^S_y, p^A_z),
\end{equation}
that describes amplitudes on orbitals of Mo and S atoms which have a dominant contribution to the orbital composition of the wave function in the bands near the energy gap.
\textit{S} and \textit{A} are symmetric $\frac{1}{\sqrt{2}}(p^t_x + p^b_x), \frac{1}{\sqrt{2}}(p^t_y + p^b_y)$ and antisymmetric: $\frac{1}{\sqrt{2}}(p^t_z - p^b_z)$ combinations of \textit{p} orbitals of S atoms.

The on-site terms of the Hamiltonian are
\begin{equation}
\epsilon_M =
\begin{pmatrix}
\Delta_0 & 0 & 0\\ 
0 & \Delta_2 & -is\lambda_M\\ 
0 & is\lambda_M & \Delta_2
\end{pmatrix},
\label{M_onsite}
\end{equation}
and
\begin{equation}
\epsilon_S =
\begin{pmatrix}
\Delta_p + V_{pp\pi} & -is\frac{\lambda_S}{2} & 0\\ 
is\frac{\lambda_S}{2} & \Delta_p + V_{pp\pi} & 0\\ 
0 & 0 & \Delta_z - V_{pp\sigma}
\end{pmatrix},
\label{S_onsite}
\end{equation}
where $s$ is 1 and $-1$ for spin up and spin down components, respectively.
The hopping matrices are introduced following Refs.~\onlinecite{silva-guillen_electronic_2016-1, khoeini_peculiar_2016} and are given in Appendix A. Slater-Koster parameters used in the implementation of the model are taken from fitting to \textit{ab initio} calculations performed in Ref.~\onlinecite{silva-guillen_electronic_2016-1}. They are presented together with the spin-orbit coupling (SOC) parameters from Ref.~\onlinecite{kosmider_large_2013} in Table~\ref{tab:params}.

\begin{table}[]
\begin{tabular}{|lll|}
\hline
\multirow{2}{*}{SOC}            & $\lambda_M$    & 0.086  \\
                                & $\lambda_S$    & 0.052  \\ \hline
\multirow{5}{*}{Crystal Fields} & $\Delta_0$     & -1.094 \\
                                & $\Delta_2$     & -1.511 \\
                                & $\Delta_p$     & -3.559 \\
                                & $\Delta_z$     & -6.886 \\ \hline
\multirow{2}{*}{M-S}            & $V_{pd\sigma}$ & 3.689  \\
                                & $V_{pd\sigma}$ & -1.241 \\ \hline
\multirow{3}{*}{M-M}            & $V_{dd\sigma}$ & -0.895 \\
                                & $V_{dd\pi}$    & 0.252  \\
                                & $V_{dd\delta}$ & 0.228  \\ \hline
\multirow{2}{*}{S-S}            & $V_{pp\sigma}$ & 1.225  \\
                                & $V_{pp\pi}$    & -0.467 \\ \hline
\end{tabular}
\caption{SOC parameters~\cite{kosmider_large_2013} and Slater-Koster parameters~\cite{silva-guillen_electronic_2016-1} used in the implementation of TB model given in eV.}
\label{tab:params}
\end{table}

\subsection{Considered system}
\begin{figure}[ht!]
\center
\includegraphics[width = 9cm]{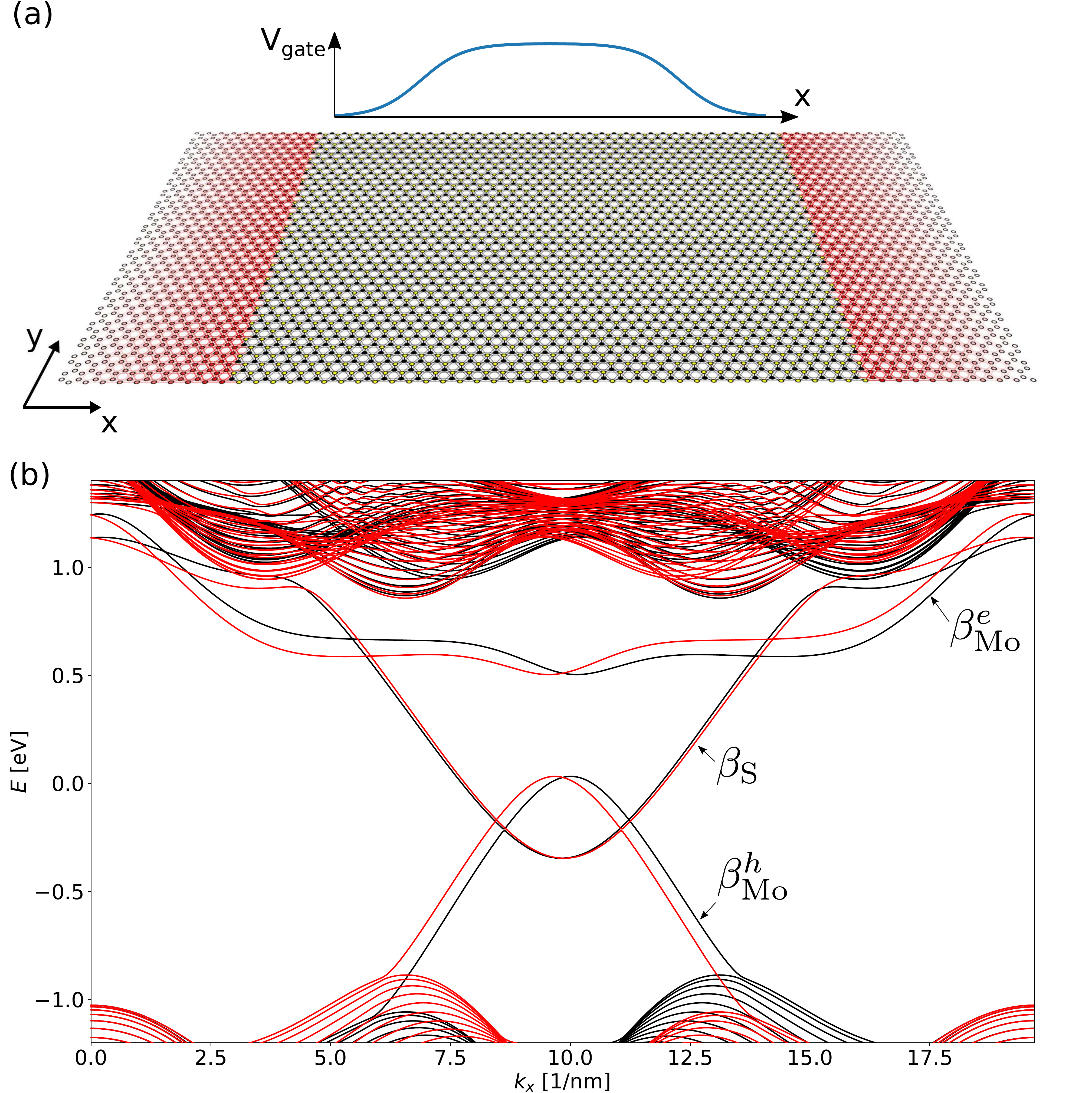}
\caption{(a) The considered $\mathrm{MoS_2}$ zigzag nanoribbon connected to two semi-infinite leads (red sites). The nonuniform potential along the $x$ direction is plotted with the blue curve in the top of the panel. (b) Dispersion relation of a 10.8 nm wide pristine ribbon. The black and red curves correspond to the spin-up and the spin-down bands, respectively. With $\beta_{\mathrm{S}}$, $\beta_{\mathrm{Mo}}^{e/h}$ we denote bands corresponding to the modes localized on sulfur and molybdenum terminated edges, respectively.}
\label{system}
\end{figure}

Figure~\ref{system}(a) presents the considered system which consists of a 10.8 nm wide $\mathrm{MoS_2}$ nanoribbon connected to semi-infinite leads. The ribbon lattice is build from S and Mo sites depicted with yellow and black circles, respectively. The wire is terminated by inequivalent atoms in the transverse ($y$) direction that define zigzag edges and is connected to two semi-infinite leads denoted with the red circles. 

\begin{figure*}[ht!]
\center
\includegraphics[width = 18cm]{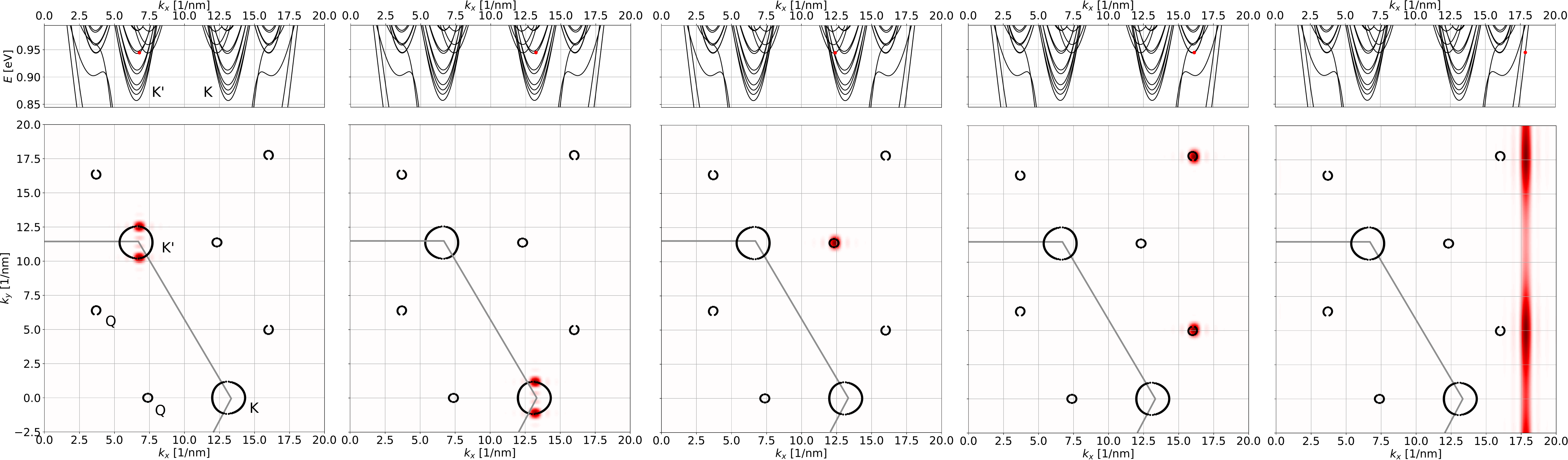}
\caption{Top panels: the dispersion relation close to the bottom of the conduction band. Bottom panels: the black dots depict Fermi contours of a bulk $\mathrm{MoS_2}$ monolayer at the Fermi energy of 0.9445 eV. The red color map that overlays the Fermi contours corresponds to the probability density of the current carrying modes denoted with the red dots in the corresponding top panels.}
\label{reciprocal_density}
\end{figure*}

In this work we are mainly interested in the structures with the broken translation symmetry. We consider a local potential variation in the ribbon that experimentally can be realized by locally gating the wire. We model the potential created in the structure as a smooth potential step by including additional position-dependent on-site elements on the diagonals of Eqs. (\ref{M_onsite}) and (\ref{S_onsite}),
\begin{equation}
V_{gate} = \frac{V_g}{2}\left[ \tanh \frac{x+L_b/2}{\lambda} - \tanh \frac{x-L_b/2}{\lambda} \right],
\label{potential}
\end{equation}
where $L_b$ is the length of the barrier and $\lambda$ sets its smoothness. For the considered model we take $\lambda=1$ nm and $L_b=6$ nm unless stated otherwise. The potential profile is presented in Fig.~\ref{system}(a).

We consider the electronic transport in the linear response regime and solve the scattering problem using the wave-function matching approach implemented in Kwant package~\cite{groth_kwant:_2014}. The code that implements the tight-binding model used for the calculation is available in Ref.~\onlinecite{d._gut_code_nodate}.

\section{Results}
\label{sec:results}
\subsection{Non-gated ribbon}
\subsubsection{Valley-polarization of current-carrying modes}
\label{sec:val_pol_modes}
We start by considering a pristine, 10 nm long, $\mathrm{MoS_2}$ zigzag ribbon. Figure \ref{system}(b) presents the dispersion relation of the wire with the black (red) curves corresponding to the spin-up (spin-down) bands. The spin-opposite bands are split in $k$ space due to the strong SOC resulting from $d$ orbitals of the heavy metal atoms. This is most clearly seen for the valence bands with the energies close to or less than $-1$ eV. The splitting is much less pronounced in the conduction bands that are present for energies exceeding 0.85 eV. In between the conduction and the valence bands there are six others bands (denoted with $\beta^h_{\mathrm{Mo}}$, $\beta_{\mathrm{S}}$, and $\beta^e_{\mathrm{Mo}}$) making the structure metallic, with no energy gap.

Let us focus on the Fermi-energy range near the conduction band minimum. In the top panels of Fig.~\ref{reciprocal_density} with the black curves we present zoom-ins of the dispersion relation of the ribbon. In the bottom panels with black points we depict the Fermi contours for an infinite monolayer $\mathrm{MoS_2}$ flake at the energy of 0.9445 eV for the positive values of the wave vector. The calculation of the Fermi surface is performed by transforming the Hamiltonian Eq.~(\ref{ham_eq}) into the momentum space and solving the resulting eigenproblem for $k_x$ and $k_y$. The bigger circles in the map form around $K$, $K'$ points, while the smaller ones are located around $Q$ points. The rings around the high-symmetry $K$ and $K'$ points are almost perfectly circular as the trigonal warping in the conduction band is much weaker than in the valence band~\cite{kormanyos_monolayer_2013}.

In each column of Fig.~\ref{reciprocal_density} we demonstrate properties of a single current-carrying mode $j$ with the wave vector value denoted with the red dot in the corresponding top panel. We calculate the wave function in the scattering region $\psi_j^l(x,y)$ by solving the scattering problem for the incoming electron in the $j$'th mode. By projecting $\psi_j^l(x,y)$ onto a basis of plane waves, we calculate the probability density in the reciprocal space at the position $(k_x, k_y)$ as
\begin{equation}
\rho_j(k_x, k_y) = \sum_{l} |\int \left[ \psi^l_j(x,y) \right]^*e^{ik_xx+ik_yy}dxdy|^2,
\label{rho_eq}
\end{equation}
where the summation is carried over orbital and spin degrees of freedom ($l$).
The resulting probability distributions are plotted on the red color map in the bottom panels of Fig.~\ref{reciprocal_density}. In the two first columns we consider states that belong to the two sets of parabolas that start at the conduction band minimum. By inspecting the positions of nonzero values of the probability density in the Brillouin zone depicted in the first (second) bottom panel, we observe that they arrange on top of the circular contours around $K'$ ($K$) points. On the other hand, the nonzero probability density for states belonging to the bands that appear at the higher energies (see the third and fourth column in Fig.~\ref{reciprocal_density}) is close to one of the $Q$ valley minima. 

We see that the valley polarization of the charge carriers in a bulk material is preserved in a semi-infinite wire and it is translated into the valley polarization of the ribbon modes. The valley polarized modes are gathered in several sets of approximately parabolic bands with the minima at different wave vectors.

Finally, in the bottom panel of the last column, we show that the density in the reciprocal space does not match any point in the Fermi contours of the bulk monolayer. This density corresponds to one of the edge modes which due to the localization on the border of the wire is delocalized in $k_y$ space accordingly to the uncertainty principle and hence lacks the valley polarization.

\begin{figure}[ht!]
\center
\includegraphics[width = 8cm]{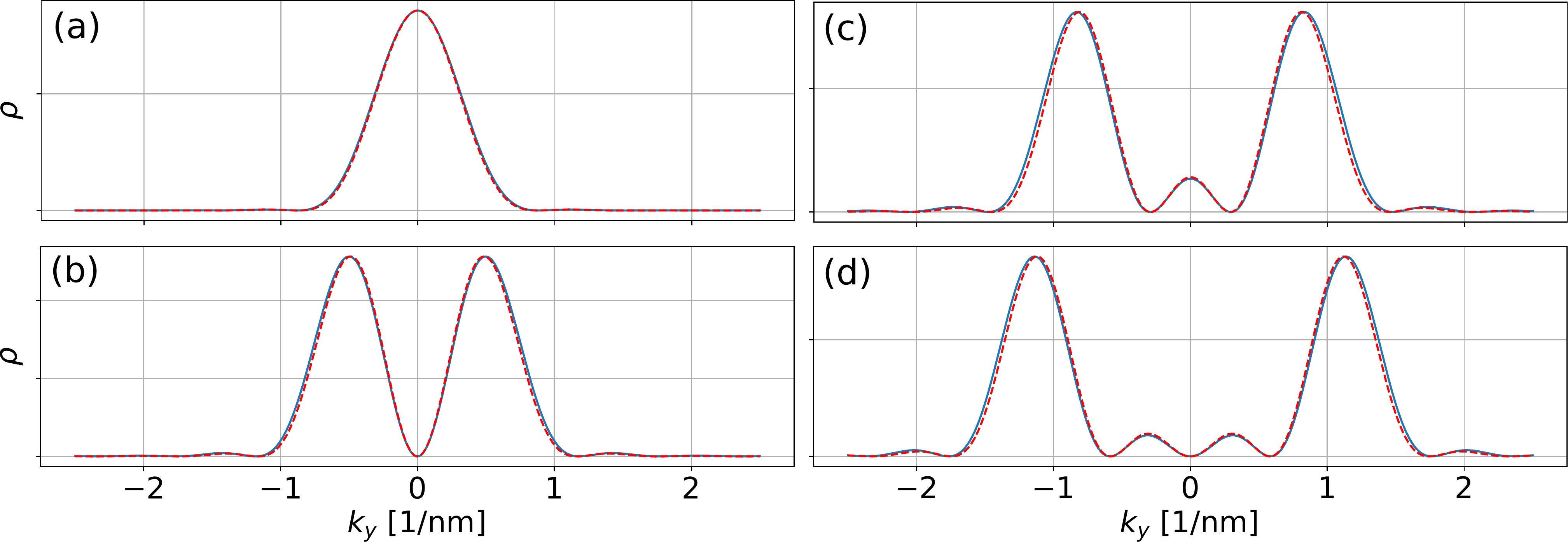}
\caption{Blue curves depict cross sections of the probability density in the reciprocal space calculated for $K$ modes. Red-dashed curves show probability density of eigenstates of an infinite quantum well of the same width as the ribbon.}
\label{crossections}
\end{figure}

Interestingly, the densities in the maps presented in Fig.~\ref{reciprocal_density} for the $K$ and $K'$ modes are distributed as a sets of dots that lie within the circular Fermi contours. To analyze further this observation let us focus on the $K$-polarized modes and inspect the cross section of the probability density in the reciprocal space for positive values of $k_x$. We consider the modes of the transverse quantization starting from the ground state.
 
\begin{figure*}[ht!]
\center
\includegraphics[width = 18cm]{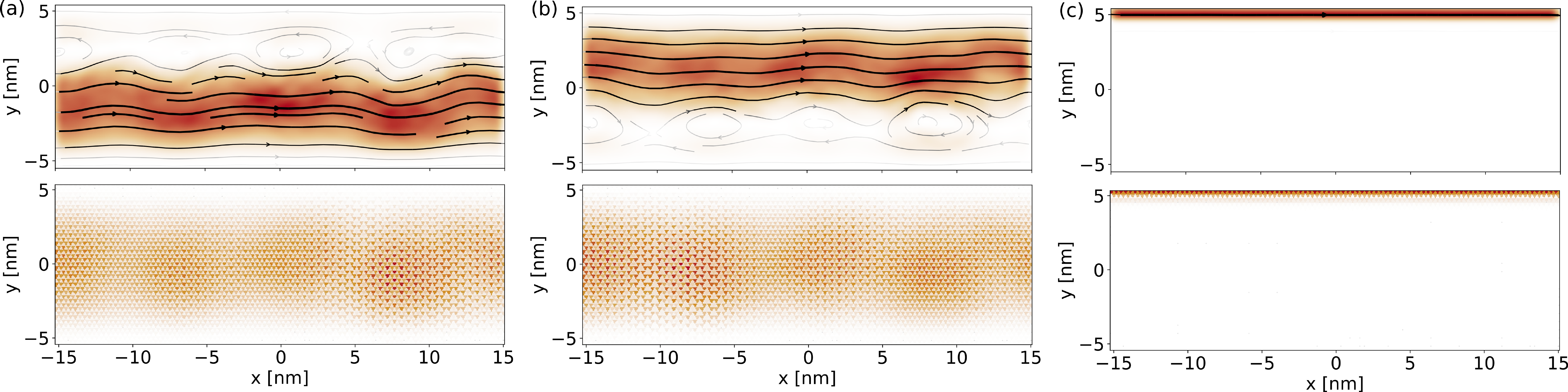}
\caption{Top panels: the probability current obtained for the ground state of the transverse quantization of $K'$ (a), $K$ (b), and $\beta^{e}_{Mo}$ edge (c) modes. Bottom panels: the corresponding probability densities.}
\label{vhe}
\end{figure*} 
 
For each mode we obtain the corresponding $k_x$ value at the considered Fermi energy and plot the probability distribution along $k_y$ with the blue curves in Fig.~\ref{crossections}. We consider only one of the spin components, and the densities for states with increasing quantization number, ranging from 1 to 4, are presented in the subsequent subplots from (a) to (d). We see that, as the quantization number increases, the number of maxima in the distribution rises. The shape of the probability distribution can be understood by considering a transform of a standing wave in an infinite quantum well of width $W$ into the momentum space. The corresponding formula for the probability density reads
\begin{equation}
\rho_n(k)=\frac{W}{\pi\hbar}\left( \frac{n\pi}{n\pi+kW}\right)^2 \mathrm{sinc}^2[1/2(n\pi-kW)],
\label{analytical_reciprocal_density}
\end{equation}
where $n$ is the integer index of the quantization number. We plot the resulting densities with red-dashed curves in Fig.~\ref{crossections}. We observe that, as $n$ increases, the number of maxima in the probability density distribution grows and the curves given by Eq.~(\ref{analytical_reciprocal_density}) overlay almost perfectly (up to a constant normalization factor) the distributions obtained from the numerical calculations. The distributions for the excited states have two main maxima that localize at the circular Fermi contours of the bulk material and are accompanied by a number of smaller local maxima that lie within these contours as showed in the bottom panels of Fig.~\ref{reciprocal_density}.

\subsubsection{Valley Hall effect in an open ribbon}
In a 2D TMDC monolayer the time reversal symmetry guarantees that the Berry curvature flips sign between $K$ and $K'$ valleys, i.e., $\mathbf{\Omega}(K)=-\mathbf{\Omega}(K')$. This leads to the valley Hall effect where the propagating electrons acquire velocity $\mathbf{v_{\bot}} = -\frac{e}{\hbar \sigma} \mathbf{J} \times \mathbf{\Omega}(K)$ perpendicular to the direction of the applied electric field that induces the current flow 
($\sigma$ stands for conductivity). Very recently signatures of the valley separation of the injected current~\cite{hung_direct_2019, wu_intrinsic_2019} were measured in the microscopic $\mathrm{MoS_2}$ Hall bar devices which followed similar observation for bilayer graphene~\cite{shimazaki_generation_2015}.

In the present system we find the valley polarization in a nanometer-size {\it semi-infinite} wire. The question remains, is there a reminiscence of the valley Hall effect in the ribbon?

For a pristine wire the translation symmetry ensures the invariance along the $x$ direction and prevents the current from drifting in the transverse direction. Let us then break the translation invariance by introducing a random on-site potential chosen from the range $[-25,25]$ meV. In top panels of Fig.~\ref{vhe} we plot the probability current carried by $K'$ (a), $K$ (b), and by the edge mode (c) at the energy of 0.87 eV for a $30$ nm long ribbon. Surprisingly, we find that the probability density currents of the $K'$ and $K$ modes show breaking of the inversion symmetry along the $y=0$ axis while the charge distribution preserves this symmetry to a good approximation. This can be understood considering that the modes are localized in different valleys that have opposite Berry curvatures which results in opposite $\mathbf{v_{\bot}}$. The broken translation symmetry now allows the electrons to scatter in the transverse direction. When the electrons are injected from the (pristine) left contact the modes polarized in $K'$ are scattered preferably towards positive $y$ values due to the positive $\mathrm{v}_{y}$ ($\Omega_z(K')<0$). Therefore, we observe current circulating close to the top edge which results in nearly zero net current along the $x$ direction. The situation is opposite for the $K$ polarized mode with the current depicted in (b) panel as $\Omega_z(K)=-\Omega_z(K')$. An analogous effect is obtained for a pristine ribbon, however there the current asymmetry is accompanied by the current circulating around primitive cells of the lattice, which preserves the translation symmetry. 

For the edge mode presented in the panel (c) we see that both the current and the charge densities are localized at the edge of the ribbon accordingly to the analysis in Sec.~\ref{sec:val_pol_modes}. 

\subsubsection{Stability of the valley polarization versus disorder}
For a bulk $\mathrm{MoS_2}$ the time-reversal symmetry protects from the intervalley scattering by nonmagnetic impurities. However, SOC is weak in the conduction band. As a result, when the Fermi energy lies above the minima of the two spin-opposite bands at $K'$ and $K$ points, the scattering between bands of the same spin and the opposite valley is allowed. It was already shown for graphene nanoribbons that the intervalley scattering is made possible by the presence of short-range scatterers~\cite{wakabayashi_perfectly_2007, wakabayashi_electronic_2009, lima_effects_2012}. Let us now investigate how disorder in the form of atomic vacancies in the atomic lattice, acting as short-range scatters, affect the valley polarization of the current and compare it to the bare electronic transport characterized by the mean free path.


\begin{figure}[ht!]
\center
\includegraphics[width = 8cm]{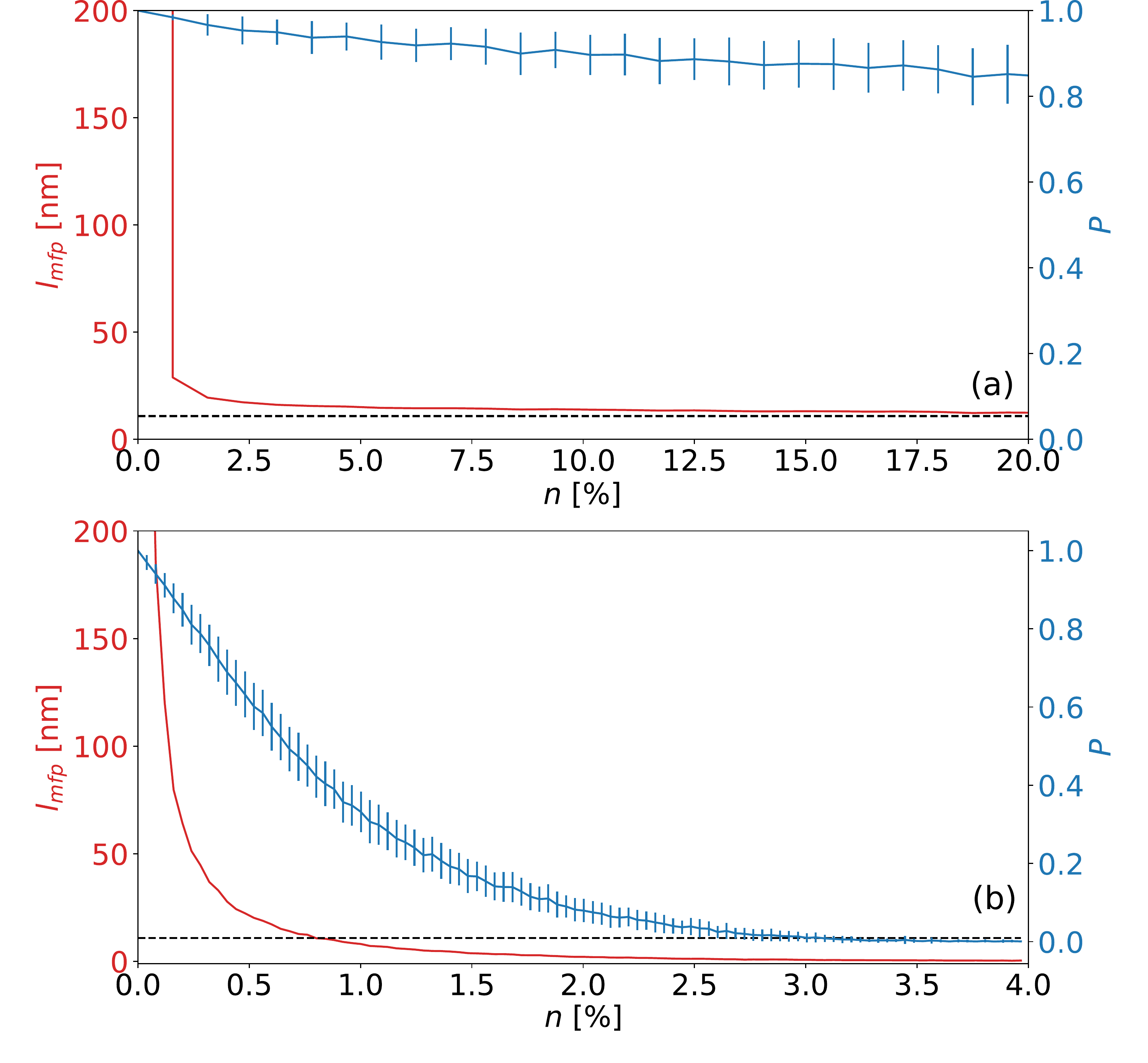}
\caption{The mean free path (red curve) and the valley polarization (blue curve) versus the abundance of vacancies calculated for disorder in the edge (a) and in the whole ribbon (b). The black dashed lines, with respect to the left axis, denote the length of the ribbon.}
\label{lP}
\end{figure}

Let us first inspect the mean free path which we calculate according to the formula~\cite{beenakker_random-matrix_1997}, $l_{\mathrm{mfp}} = 2L \left\langle T_{tot}\right\rangle /\pi(N-\left\langle T_{tot}\right\rangle)$, where $N$ is the number of current carrying modes at a given Fermi energy, $L$ is the length of the considered scattering region (here 10 nm), and $\left\langle T_{tot} \right\rangle$ is the total transmission  averaged over 200 realizations of the disorder at the Fermi energy of 0.9 eV .

The resulting mean free path versus the vacancy abundance is plotted in Fig.~\ref{lP} with the red curves. In panel (a) we study a ribbon with disordered edge, i.e., we include vacancies only on the two atomic rows closest to each edge. We observe that even a few vacancies on the edge---that break the zigzag structure---significantly decrease the mean free path by blocking of the transport through the edge.~\cite{ridolfi_electronic_2017} $l_{\mathrm{mfp}}$ saturates at the value close to the ribbon length as $l_{\mathrm{mfp}}/L\simeq(N-4)/2\pi\simeq1$ (where 4 stands for the four, blocked, edge modes and $N = 12$).

We now estimate the valley polarization of the current as:
\begin{equation}
P = \frac{\left\langle T_{K\rightarrow K} \right\rangle+ \left\langle T_{K' \rightarrow K'}\right\rangle}{N_{v}},
\end{equation}
where $N_{v}$ is the number of the valley polarized modes at the input of the ribbon and $\langle T_{K^{(')} \rightarrow K^{(')}} \rangle$ is an averaged transmission probability from $K$ ($K'$) to $K$ ($K'$) modes. For the considered Fermi energy we identify the valley polarization of the modes by inspecting the wave vector corresponding to each mode. For the disordered edge, $P$ drops only slightly when the number of vacancies is increased [Fig.~\ref{lP}(a)] as the bulk modes are weakly influenced by the vacancies on the edge of the ribbon.

Now let us turn our attention to the vacancies present in the whole ribbon. In Fig.~\ref{lP}(b) we observe that when their number is increased from zero to $0.5\%$ there is a rapid drop of $l_\mathrm{mfp}$. This can be attributed to the Anderson localization~\cite{anderson_absence_1958, kleftogiannis_conductance_2013, colomes_antichiral_2018}.  At around $n=0.8\%$ $l_{\mathrm{mfp}}$ becomes smaller than the considered ribbon length [dashed line in Fig.~\ref{lP}(b)] and the transport becomes diffusive. Finally, as the number of vacancies increases further, $l_{\mathrm{mfp}}$ saturates at the value of 0.4 nm.

For a moderate vacancy concentration $P$ drops, in a good approximation, linearly with $n$. As a result, even for a ribbon located deep in the ballistic regime, a moderate concentration is sufficient to distort the perfect valley polarization. At the transition region separating the ballistic and the diffusive transport, when $l_{\mathrm{mfp}} \simeq L$, $P \simeq 0.4$, which is a signature of the strong valley mixing by the scatterers. As a result, we expect that the valley polarization of the current can be maintained only in the pristine samples where the vacancy concentration is less than 0.1 \%. This is satisfied in $\mathrm{MoS_2}$ samples obtained by mechanical exfoliation or by chemical vapor deposition for which the reported defects density is $3.5 \times 10^{13}\; \mathrm{cm}^{-2}$.~\cite{hong_exploring_2015}

\subsection{Gated nanoribbon}
Very recently, it became possible to create locally gated $\mathrm{MoS_2}$ nanostructures suitable for experimental quantum transport measurements~\cite{kim_genuine_2019, zheng_patterning_2019}. In this section we study a model system of a locally gated ribbon, i.e. a zigzag nanoribbon with a potential step along the $x$ direction. 

\begin{figure}[ht!]
\center
\includegraphics[width = 7.5cm]{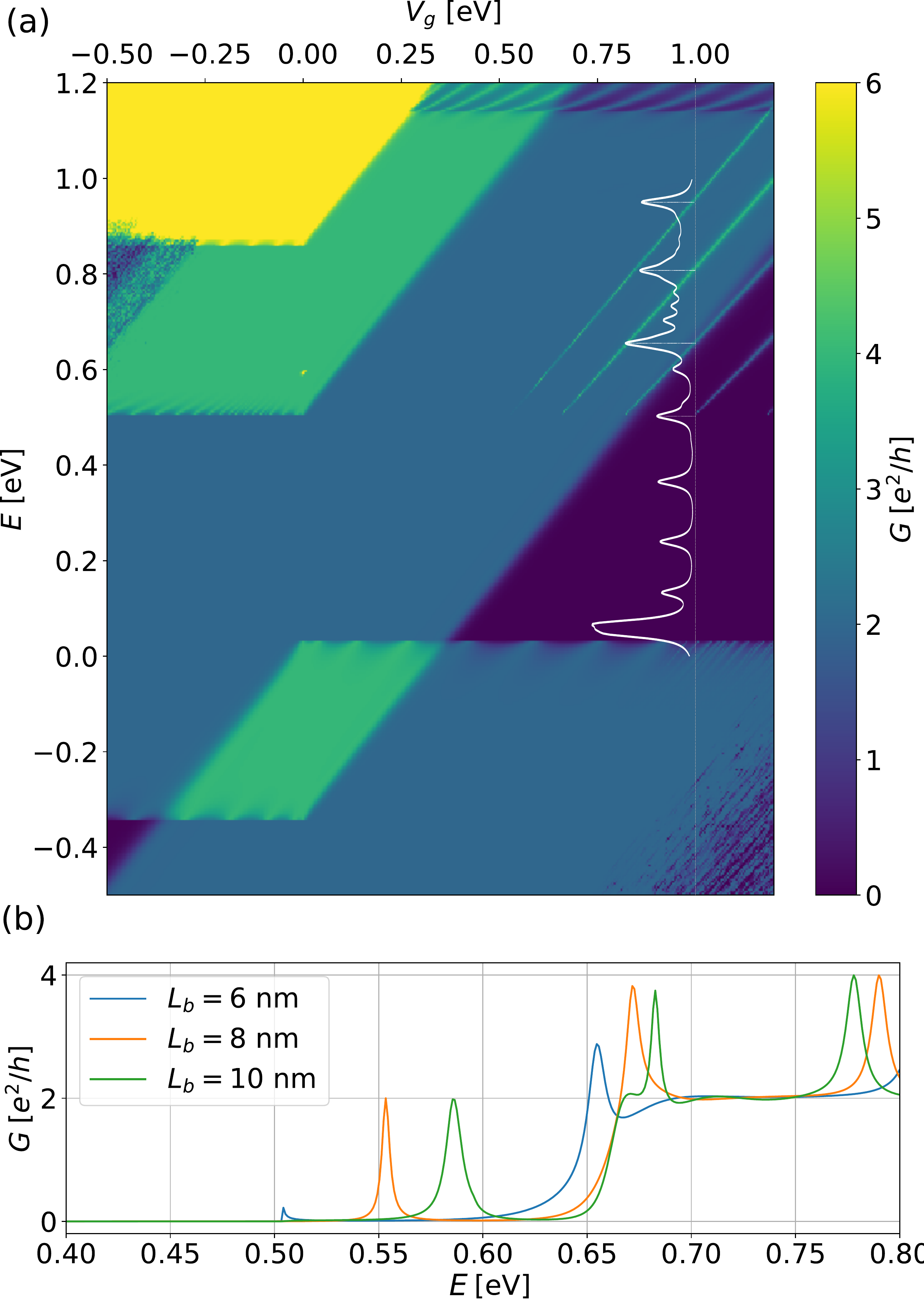}
\caption{(a) Conductance map of a ribbon with a potential step as a function of the Fermi energy and the magnitude of the local potential. The white curve presents LDOS calculated in the potential barrier with peaks due to bound states originating from quasihole edge states. (b) The conductance cross section for $V_g=1$ eV as a function of the the Fermi energy for three values of the barrier length.}
\label{G_map}
\end{figure}

The conductance as a function of the Fermi energy and $V_g$ that controls the height of the potential step in the ribbon is presented in Fig.~\ref{G_map}(a). Let us first focus on the case of a pristine ribbon with $V_g = 0$. In this case the conductance quantized in multiples of $2e^2/h$ is proportional to the number of current carrying modes at a given Fermi energy. For $E < 0.9$ eV the current is carried by the edge modes and we find stable plateaus of conductance. Such a stable conductance quantization was recently measured in $\mathrm{WTe_2}$ nanodevices where it was a signature of the quantum spin Hall effect~\cite{wu_observation_2018}---the two spin-opposite modes carried the topologically protected current on the opposite edges of the sample. Here however, we find that each edge band is Kramer degenerate and consists of a doublet of counterpropagating modes for each spin polarization---see the bands denoted with $\beta_{\mathrm{Mo}}^{h}$, $\beta_{\mathrm{S}}$, $\beta_{\mathrm{Mo}}^{e}$ in Fig.~\ref{system}(b). Therefore, even though we find a conductance $2e^2/h$, it results from the transport of spin opposite modes located at the same edge.

This becomes clear when we consider nonzero values of $V_g$ that allow us to gate out the current. In the map we find a dark-blue triangle for which $G=0$. In this parameter regime, in the nongated part of the nanostructure, the current is carried through $\beta_{\mathrm{S}}$ modes, while, in the gated part, the chemical potential shifts the bands such at the Fermi level there is only $\beta_{\mathrm{Mo}}^{h}$ band available. As a result, for a finite width sample those edge current-carrying states are spatially disconnected and the transport through the ribbon is blocked despite its metallic character. 

\subsubsection{Resonant transport}
In the map in Fig.~\ref{G_map}(a) for $E > 0.5$ eV and $V_g > 0.5$ eV we observe a set of sharp lines of increased conductance. For these values of the Fermi energy, in the region of the ribbon far from the potential barrier, the current is carried by both the edge modes $\beta_{\mathrm{S}}$ and $\beta_{\mathrm{Mo}}^{e}$ located at the sulfur and molybdenum terminated [see Fig.~\ref{vhe}(c)] edges, respectively.

The variation of the chemical potential along the ribbon, due to the gate potential, creates effectively an n-p-n junction for the modes located on the Mo-terminated edge. Here n (p) stands for the regime where the current is carried through electronlike $\beta_{\mathrm{Mo}}^{e}$ (holelike $\beta_{\mathrm{Mo}}^{h}$) band. The n and p parts are separated by the regions where the Fermi energy is located in the energy gap between the $\beta_{\mathrm{Mo}}^{h}$ and $\beta_{\mathrm{Mo}}^{e}$ bands. As a result, the constructive interference of the states from the $\beta_{\mathrm{Mo}}^{h}$ band in the potential barrier gives rise to a series of localized bound states. We confirm this by calculating local density of states for a system with disconnected leads and integrating it over the barrier region,
\begin{equation}
DOS_{\mathrm{barrier}} = \sum_i \int_{-W/2}^{W/2}\int^{L_b/3}_{-L_b/3} |\Psi_i|^2f_L(E, E_i, \Gamma) dx dy,
\end{equation}
where $f_L$ is a Lorentzian function with the width parameter $\Gamma=0.02$ eV.
The calculated density for $V_g=1$ eV is plotted with the white curve overlaying the conductance map in Fig.~\ref{G_map}(a). We observe maxima in the density of states that correspond to the edge states localized in the barrier region. They appear at the energies that correspond to the resonance energies, evidencing that the conductance increases when the incoming electron energy matches the energy of one of the  bound states. Similar resonances on a n-p-n junction were observed in bilayer graphene~\cite{eich_spin_2018-1} but here they are rather due to the localized edge states.

Another important feature of those states is their energy distribution. Note that the quasihole edge band, out of which the bound states form, far from its maximum is approximately linear [see the spin-split bands that start around zero energy and are oriented towards valence bands in Fig.~\ref{system}(b)]. The condition for the formation of the bound states is the quantization of the wave vector $k=n\pi/W$ where $n$ is an integer. As a result of the linear dependence of $E$ on $k$ for the quasihole band, the bound states are equidistant in energy. Accordingly, when we increase $L_b$ we observe that the spacing between the resonances also rises as expected for bound states -- see Fig.~\ref{G_map}(b).

Note that the resonances are observed only for the Fermi energies for which the electrons are injected in a band localized on the same edge as the edge on which the bound states are located. As a result the resonances vanish below $E=0.5$ eV, for which in the non-gated part the Fermi energy is located below $\beta_{\mathrm{Mo}}^{e}$ band. Despite this, we still find localized states in the gated region in this energy range [see the LDOS peaks in the white curve in Fig.~\ref{G_map}(a)]. 

\subsubsection{Gate induced valley polarization of the current}
\begin{figure}[ht!]
\center
\includegraphics[width = 8cm]{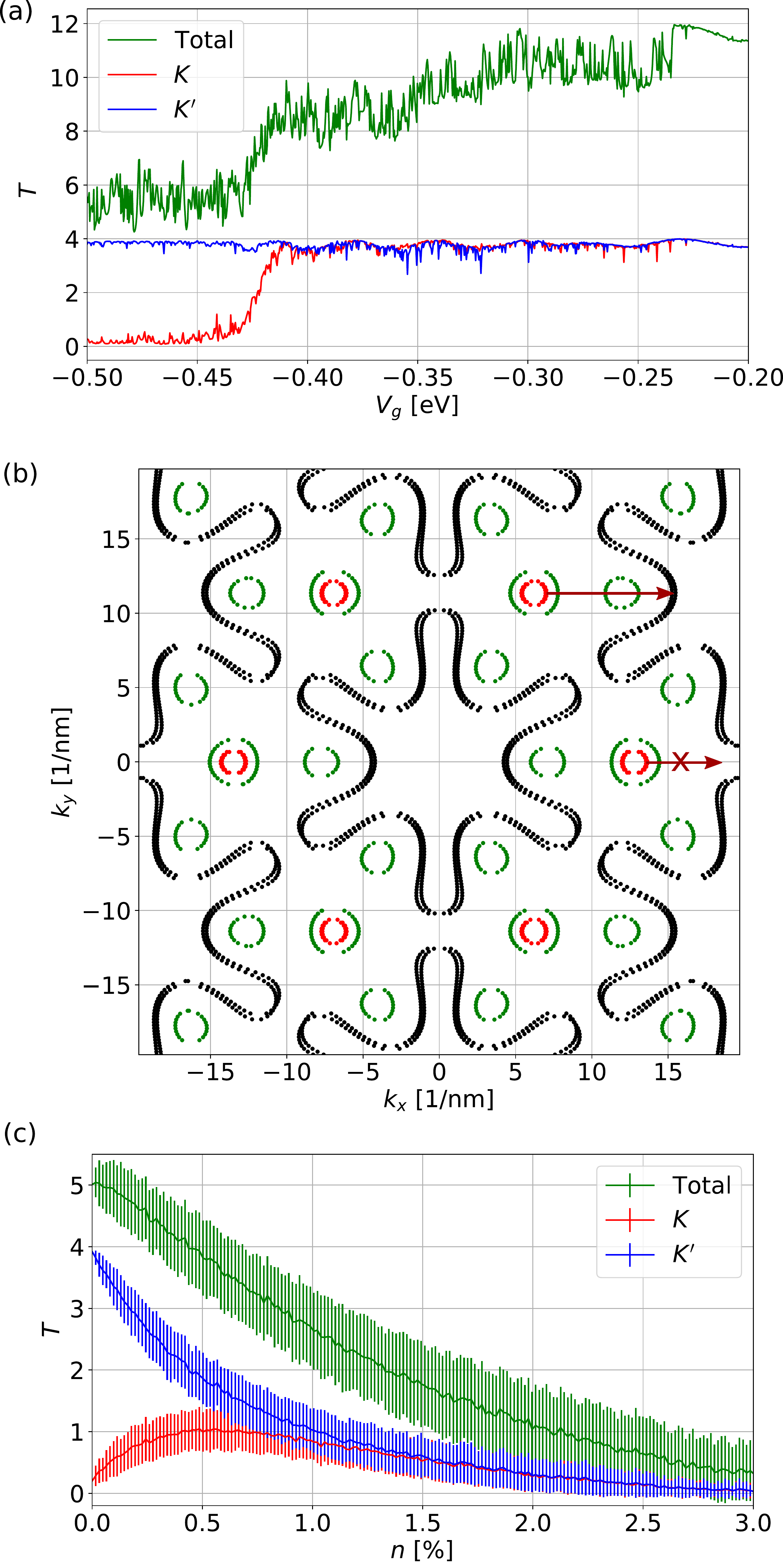}
\caption{(a) The total transmission probability (green curve) for the Fermi energy $E=0.9$ eV and the sums of transmission probabilities of $K$ (red curve) and $K'$ (blue curve) modes at the output of the locally gated ribbon. (b) Fermi contours at $E=0.9$ eV (red), $E=1$ eV (green), and $E=1.35$ eV (black). The arrows present $k_x$ scattering processes between the bands with a positive group velocity for an electron in the $K'$ valley (straight arrow) and a forbidden process for an electron in the $K$ valley (crossed arrow). (c) Total transmission (green) and the transmission of $K$ (red) and $K'$ polarized modes for $V_g = -0.45$ eV in the presence of vacancies.}
\label{barrier_valley_polarization}
\end{figure}

Let us now focus on the transport through the ribbon when the Fermi energy lies above the conduction band minimum. Namely, we consider $E = 0.9$ eV and calculate the conductance for negative values of the local gate potential. The result is presented in Fig.~\ref{barrier_valley_polarization}(a) with the green curve which exhibits a sudden drop when the local gate potential reaches $-0.42$ eV.

\begin{table}[]
\begin{tabular}{ |c|c|c|c| } 
\hline
Mode type & $K'$ & $K$ & edge\\
\hline
\multirow{4}{10em}{Transmission probability} & 0.991 & 0.025 & 0.103\\ 
& 0.986 & 0.016 & 0.114\\ 
& 0.976 & 0.089 & 0.337\\ 
& 0.968 & 0.070 & 0.333\\ 
\hline
\end{tabular}
\caption{Transmission probabilities to the $K'$, $K$ and edge modes for $V_g=-0.45$ eV.}
\label{transmissions}
\end{table}

We calculate the transmission probability {\it to} the $K$ and $K'$ modes at the right semi-infinite lead. The results are plotted with the red and blue curves in Fig.~\ref{barrier_valley_polarization}(a) respectively, and the actual values of the transmission coefficient of the outgoing modes can be found in Table~\ref{transmissions} for a single value of $V_g$. We observe that the conductance drop follows from the almost completely blocked transport of $K$ modes through the gated region. This results in a valley polarization of the current flowing from the structure with the polarization ratio being on average $P_v = |(T_{K} - T_{K'})|/T_{tot}=75\%$.

\begin{figure}[ht!]
\center
\includegraphics[width = 8cm]{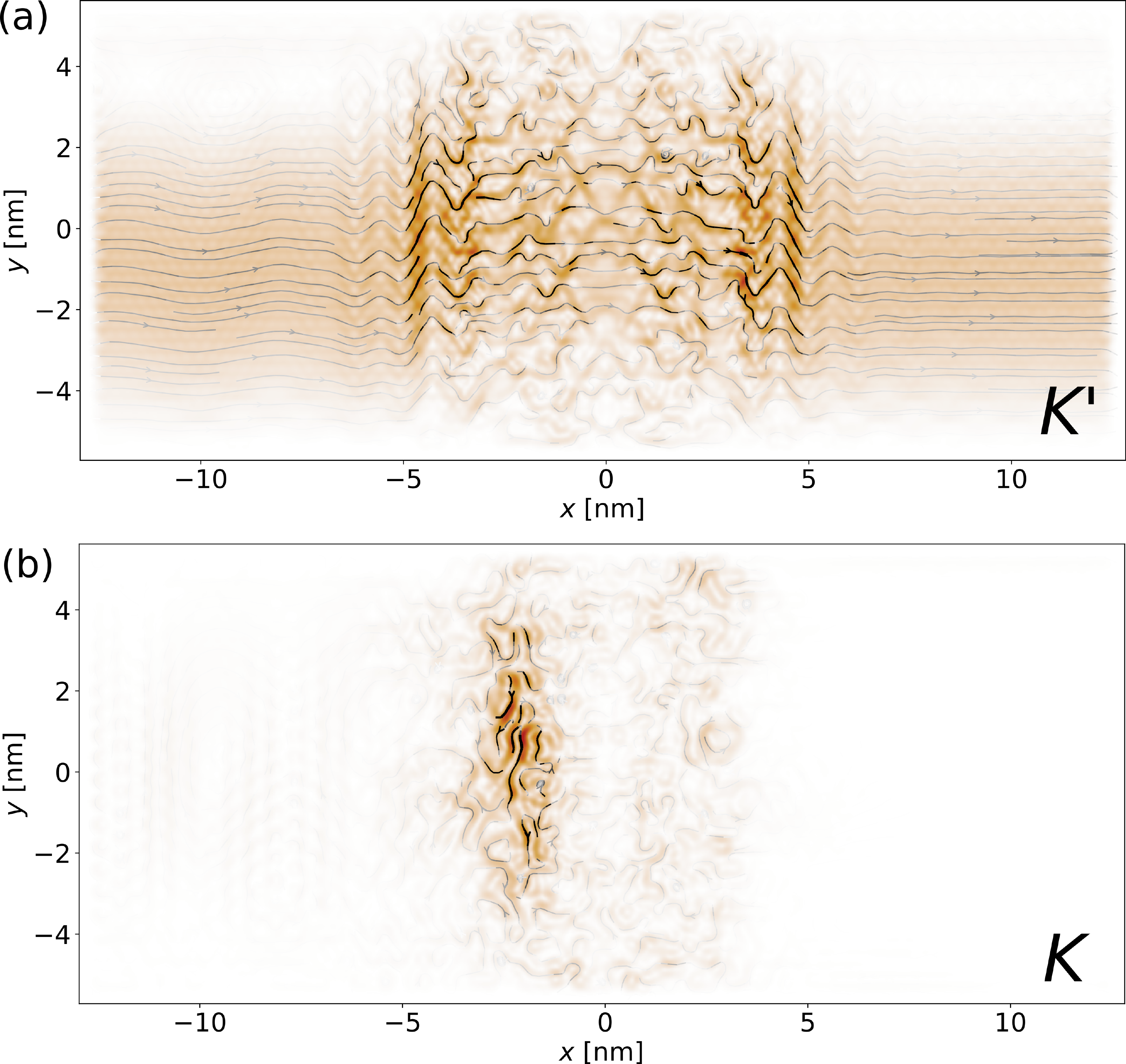}
\caption{(a) The probability current for an electron injected in a $K'$ valley polarized mode with a nonzero current distribution at the right lead. (b) The probability current for an electron injected in a $K$ valley polarized mode that is reflected at the potential well.}
\label{barrier_current}
\end{figure}

The effect of the valley polarization can be understood by comparing the band structure in the gated region and in the region distant from the potential step. The potential step in the ribbon allows for reflection of the electrons propagating along the channel, which results in the $k_x$ momentum scattering. On the other hand, the ribbon is to a good approximation homogeneous along the $y$ direction, making the scattering in $k_y$ unlikely (recall that the bulk modes have well defined $k_y$ values as can be seen in Fig.~\ref{reciprocal_density}). If we now compare the Fermi contours for $E=0.9$ eV that correspond to the nongated region in the channel with the Fermi contours for $E=1.35$ eV that correspond to the band structure in the gated region [see Fig.~\ref{barrier_valley_polarization}(b)], we observe that for the electrons injected in the $K$ modes there are no states with the positive group velocity at both energies that preserve $k_y$. On the other hand, such states do exist for the $K'$ polarized modes. Therefore, the $K'$ modes in the ribbon propagate freely through the gated region while the $K$ modes are almost completely reflected which can be observed in Fig.~\ref{barrier_current}. The small nonzero transmission probability of the $K$ modes can be assigned to a small $k_y$ momentum scattering which results from breaking of the mirror symmetry along the $y$ direction due to the presence of nonequivalent ribbon edges.

Finally, in Fig.~\ref{barrier_valley_polarization}(c) we show the transmission  of $K$ and $K'$ polarized modes with the red and blue curves, respectively, in the presence of short-range scatterers in the form of vacancies for $V_g = -0.45$ eV. The values are averaged over 200 realizations of the disorder and the vertical lines denote the standard deviation. We observe that as the vacancy abundance $n$ is increased, the total transmission [green curve in Fig.~\ref{barrier_valley_polarization}(c)] monotonically drops. Interestingly, the transmission of $K$ modes, heavily limited for the case of pristine ribbon [Fig.~\ref{barrier_valley_polarization}(a)], exhibits nonmonotonic dependence on $n$. When the vacancy abundance is increased from 0 to 0.5 \% the transmission of $K$ modes increases as the short-range scatterers induce momentum scattering, enabling transport through the gated region. When $n$ is increased above 0.5 \%, the transmission coefficients for both $K$ and $K'$ modes start to decrease, as disorder limits the transport through the ribbon regardless of the valley polarization of the modes.

\section{Summary}
\label{sec:summary}
In this paper we studied the electronic transport in a $\mathrm{MoS_2}$ nanoribbon. We characterized current carrying modes in a pristine ribbon demonstrating that, despite finite width of the wire, the wire bulk modes modes are valley polarized and belong to $K'$, $K$, and $Q$ valleys depending on their wave vector. For the $K$ and the $K'$ polarized modes we found that the probability current is shifted towards the edges of the ribbon compatible with the valley Hall effect. In a disordered ribbon---populated by atomic vacancies---we found that the valley polarization drops linearly with the increase of the vacancy concentration and can drop below 0.5 even when the transport itself is still ballistic. 
We considered a model of a gated ribbon and found that, despite the metallic nature of the zigzag wire, the current carried by the edge modes can be gated out and that in the gated region resonant quasihole edge states are formed. The latter results in sharp conductance peaks equidistant in energy due to the linear nature of the quasihole edge band. Finally, we found that by creating a negative potential step across the ribbon a nearly perfect valley polarization can be obtained due to the mismatch between the band structure near the $K$ and the $K'$ points.  

\begin{acknowledgements}
D.G., M.P., and M.P.N. acknowledge support within POIR.04.04.00-00-3FD8/17 project carried out within the HOMING programme of the Foundation for Polish Science co-financed by the European Union under the European Regional Development Fund. D.S.~was supported by CNCS-UEFISCDI, with project No. PN-III-P1-1.2-PCCDI-2017-0338. The calculations were performed on PL-Grid Infrastructure. The authors acknowledge helpful discussions with F. Pi\'echon, A.~Akhmerov, M.~Wimmer, J.~Weston, D.~Varjas, and T.~Rosdahl.
\end{acknowledgements}

\appendix
\section{Hopping matrices}
\begin{figure}[ht!]
\center
\includegraphics[width = 8cm]{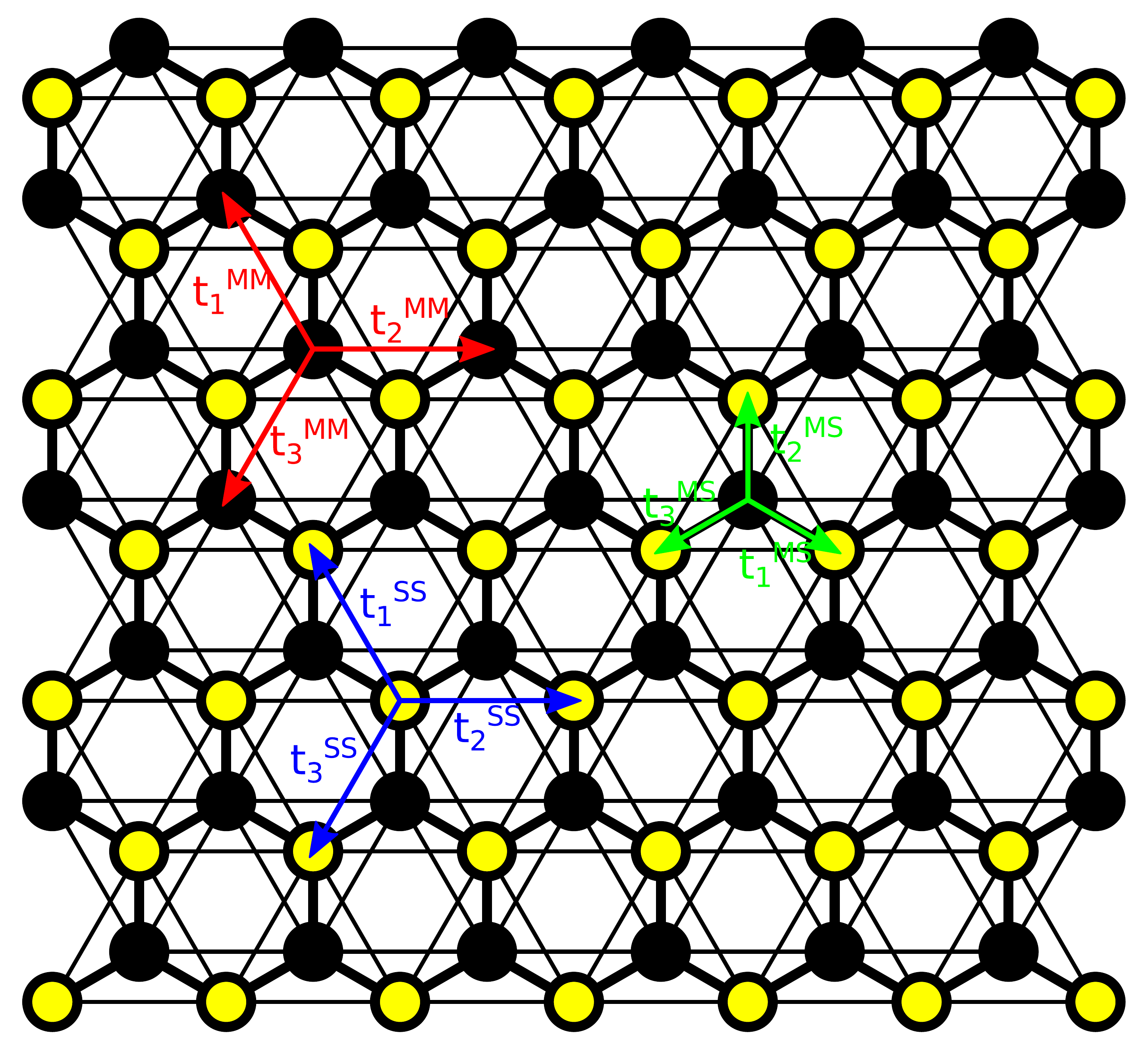}
\caption{The lattice model for the considered system with denoted intralattice (red, blue arrows) and interlattice (green) hoppings.}
\label{hoppings}
\end{figure}
The interlattice hopping matrices for the adopted TB model are as follows,
\begin{widetext}
\begin{equation}
t_1^{MS} =  \frac{\sqrt2}{7\sqrt7}
\begin{pmatrix}
-9V_{pd\pi}+\sqrt3\ V_{pd\sigma} & 3\sqrt3\ V_{pd\pi}-V_{pd\sigma} & 12V_{pd\pi}+\sqrt3\ V_{pd\sigma} \\
5\sqrt3\ V_{pd\pi}+3V_{pd\sigma} & 9V_{pd\pi}-\sqrt3\ V_{pd\sigma} & -2\sqrt3\ V_{pd\pi}+3V_{pd\sigma} \\
-V_{pd\pi}-3\sqrt3\ V_{pd\sigma} & 5\sqrt3\ V_{pd\pi}+3V_{pd\sigma} & 6V_{pd\pi}-3\sqrt3\ V_{pd\sigma}
\end{pmatrix},
\end{equation}
\begin{equation}
t_2^{MS} =  \frac{\sqrt2}{7\sqrt7}
\begin{pmatrix}
0 & -6\sqrt3V_{pd\pi}+2V_{pd\sigma} & 12V_{pd\pi}+\sqrt3V_{pd\sigma}\\
0 & -6V_{pd\pi}-4\sqrt3V_{pd\sigma} & 4\sqrt3V_{pd\pi}-6V_{pd\sigma}\\
14V_{pd\pi} & 0 & 0
\end{pmatrix},
\end{equation}
\begin{equation}
t_3^{MS} =  \frac{\sqrt2}{7\sqrt7}
\begin{pmatrix}
9V_{pd\pi}-\sqrt3V_{pd\sigma} & 3\sqrt3V_{pd\pi}-V_{pd\sigma} & 12V_{pd\pi}+\sqrt3V_{pd\sigma}\\
-5\sqrt3V_{pd\pi}-3V_{pd\sigma} & 9V_{pd\pi}-\sqrt3V_{pd\sigma} & -2\sqrt3V_{pd\pi}+3V_{pd\sigma}\\
-V_{pd\pi}-3\sqrt3V_{pd\sigma} & -5\sqrt3V_{pd\pi}-3V_{pd\sigma} & -6V_{pd\pi}+3\sqrt3V_{pd\sigma}
\end{pmatrix}.
\end{equation}
The intralattice hopping elements are
\begin{equation}
t_1^{MM} =  \frac{1}{4}
\begin{pmatrix}
3V_{dd\delta}+V_{dd\sigma} & {\frac{\sqrt3}{2}}\left(-V_{dd\delta}+V_{dd\sigma}\right) & -{\frac{3}{2}}\left(V_{dd\delta}-V_{dd\sigma}\right)\\
{\frac{\sqrt3}{2}}\left(-V_{dd\delta}+V_{dd\sigma}\right) & {\frac{1}{4}}\left(V_{dd\delta}+12V_{dd\pi}+3V_{dd\sigma}\right) & {\frac{\sqrt3}{4}}\left(V_{dd\delta}-4V_{dd\pi}+3V_{dd\sigma}\right)\\
-{\frac{3}{2}}\left(V_{dd\delta}-V_{dd\sigma}\right) & {\frac{\sqrt3}{4}}\left(V_{dd\delta}-4V_{dd\pi}+3V_{dd\sigma}\right) & {\frac{1}{4}}\left(3V_{dd\delta}+4V_{dd\pi}+9V_{dd\sigma}\right)
\end{pmatrix},
\end{equation}
\begin{equation}
t_2^{MM} =  \frac{1}{4}
\begin{pmatrix}
3V_{dd\delta}+V_{dd\sigma} & \sqrt3\left(V_{dd\delta}-V_{dd\sigma}\right) & 0\\
\sqrt3\left(V_{dd\delta}-V_{dd\sigma}\right) & V_{dd\delta}+{3V}_{dd\sigma} & 0\\
0 & 0 & 4V_{dd\pi}
\end{pmatrix},
\end{equation}
\begin{equation}
t_3^{MM} =  \frac{1}{4}
\begin{pmatrix}
3V_{dd\delta}+V_{dd\sigma} & {\frac{\sqrt3}{2}}\left(-V_{dd\delta}+V_{dd\sigma}\right) & {\frac{3}{2}}\left(V_{dd\delta}-V_{dd\sigma}\right)\\
{\frac{\sqrt3}{2}}\left(-V_{dd\delta}+V_{dd\sigma}\right) & {\frac{1}{4}}\left(V_{dd\delta}+12V_{dd\pi}+3V_{dd\sigma}\right) & -{\frac{\sqrt3}{4}}\left(V_{dd\delta}-4V_{dd\pi}+3V_{dd\sigma}\right)\\
{\frac{3}{2}}\left(V_{dd\delta}-V_{dd\sigma}\right) & -{\frac{\sqrt3}{4}}\left(V_{dd\delta}-4V_{dd\pi}+3V_{dd\sigma}\right) & {\frac{1}{4}}\left(3V_{dd\delta}+4V_{dd\pi}+9V_{dd\sigma}\right)
\end{pmatrix},
\end{equation}
\end{widetext}
\begin{equation}
t_1^{SS} =  \frac{1}{4}
\begin{pmatrix}
{3V}_{pp\pi}+V_{pp\sigma} & \sqrt3\left(V_{pp\pi}-V_{pp\sigma}\right) & 0\\
\sqrt3\left(V_{pp\pi}-V_{pp\sigma}\right) & V_{pp\pi}+{3V}_{pp\sigma} & 0\\
0 & 0 & 4V_{pp\pi}
\end{pmatrix},
\end{equation}
\begin{equation}
t_2^{SS} =  \frac{1}{4}
\begin{pmatrix}
V_{pp\sigma} & 0 & 0\\
0 & V_{pp\pi} & 0\\
0 & 0 & V_{pp\pi}
\end{pmatrix},
\end{equation}
and
\begin{equation}
t_3^{SS} =  \frac{1}{4}
\begin{pmatrix}
{3V}_{pp\pi}+V_{pp\sigma} & -\sqrt3\left(V_{pp\pi}-V_{pp\sigma}\right) & 0\\
-\sqrt3\left(V_{pp\pi}-V_{pp\sigma}\right) & V_{pp\pi}+{3V}_{pp\sigma} & 0\\
0 & 0 & 4V_{pp\pi}
\end{pmatrix}.
\end{equation}
The mapping between the hoppings and the $t_{i,j}$ elements of the Hamiltonian Eq.~\eqref{ham_eq} is presented in Fig.~\ref{hoppings}.

\bibliography{references}

\end{document}